\documentclass[12pt,a4paper]{article}
\usepackage{fullpage}
\usepackage[centertags]{amsmath}
\allowdisplaybreaks[4]
\numberwithin{equation}{section}
\usepackage{amssymb}
\usepackage{bm}
\usepackage{bbm}
\usepackage{mathrsfs}
\usepackage{float}
\usepackage{graphicx}
\usepackage{url}
\usepackage[dvips,hyperindex]{hyperref}
\usepackage{tocbibind}
\usepackage{subfigure}

\newcommand{\vet}[1]{\ensuremath{\hskip-1pt\vec{\hskip1pt#1}}}

\begin{document}

\begin{flushright}
\textsf{9 October 2006}
\\
\textsf{hep-ph/0608070}
\end{flushright}

\vspace{1cm}

\begin{center}
\large
\textbf{Neutrino Flavor States and the Quantum Theory of Neutrino Oscillations}
\normalsize
\\[0.5cm]
\large
Carlo Giunti
\normalsize
\\[0.5cm]
INFN, Sezione di Torino, and Dipartimento di Fisica Teorica,
\\
Universit\`a di Torino,
Via P. Giuria 1, I--10125 Torino, Italy
\\[0.5cm]
\begin{minipage}[t]{0.8\textwidth}
\begin{center}
\textbf{Abstract}
\end{center}
The definition and derivation of flavor neutrino states
in the framework of standard Quantum Field Theory is reviewed,
clarifying some subtle points.
It is shown that a
flavor neutrino state that describes a neutrino produced or detected in
a charged-current weak interaction process
depends on the process under consideration
and is appropriate
for the description of neutrino oscillations
as well as for the calculation of neutrino production or detection rates.
Hence, we have a consistent framework for the description of
neutrino oscillations and interactions in neutrino oscillation experiments.
The standard flavor neutrino states are obtained as
approximations which describe neutrinos in experiments that are not sensitive
to the dependence of neutrino interactions on the neutrino mass differences.
It is also shown that the oscillation probability
can be derived either through the usual
light-ray time $=$ distance approximation
or through an average of the space-time dependent oscillation probability
over the unobserved propagation time.
\end{minipage}
\\[0.5cm]
To be published as a \emph{Topical Review} in
\\
\emph{Journal of Physics G: Nuclear and Particle Physics}
\end{center}


\bigskip \hrule \bigskip
\tableofcontents
\bigskip \hrule \bigskip

\section{Introduction}
\label{Introduction}

The standard theory of neutrino oscillations
was derived in 1975-76
\cite{Eliezer:1976ja,Fritzsch:1976rz,Bilenky:1976cw,Bilenky:1976yj}
under the assumption that
a flavor neutrino $\nu_{\alpha}$,
produced or detected in a charged-current weak interaction process
together with a charged lepton with flavor $\alpha$ ($\alpha=e,\mu,\tau$)
is described by the standard flavor state
\begin{equation}
| \nu_{\alpha} \rangle
=
\sum_{k} U_{\alpha k}^* \, | \nu_{k} \rangle
\,,
\label{1331}
\end{equation}
where $U$ is the unitary mixing matrix of the neutrino fields
and $ | \nu_{k} \rangle $ are the Fock states of the massive neutrino fields,
with corresponding masses $m_{k}$
(see the reviews in
Refs.~\cite{Bilenky:1978nj,Bilenky:1987ty,hep-ph/9812360,hep-ph/0202058,hep-ph/0306239}).

In Ref.~\cite{Giunti:1992cb} it has been shown that
the flavor state in Eq.~(\ref{1331}) is \emph{not}
a quantum of the flavor field $\nu_{\alpha}$
(it is not annihilated by $\nu_{\alpha}$
if the neutrino masses are taken into account).
However,
it was shown that the flavor state in Eq.~(\ref{1331})
describes a flavor neutrino $\nu_{\alpha}$ in the realistic
ultrarelativistic approximation
($ m_{k} \ll E $, where $E$ is the neutrino energy) \cite{Giunti:1992cb}.

Later \cite{hep-ph/9501263},
it has been discovered that
it is possible to construct a Fock space of flavor states,
which allows an alternative description of
neutrino oscillations
(see also Refs.~\cite{hep-th/9803157,hep-ph/9807266,hep-ph/9907382,hep-ph/0102001,hep-th/0204184,hep-ph/0212402,hep-ph/0201188,hep-th/0408228,hep-ph/0505178}).
However,
this construction suffers from mysterious features
(the number of flavor  Fock spaces is infinite, depending on arbitrary mass parameters;
each flavor vacuum and the corresponding ladder operators are time-dependent;
the ladder operators at different times do not satisfy the canonical
anticommutation relations)
and
it has been shown that the flavor Fock states
cannot be applied to the calculation of interaction processes
\cite{hep-ph/0312256,hep-ph/0604069}.

Let us emphasize that the inapplicability of
flavor Fock space theories to the calculation of neutrino interactions
is a crucial shortcoming for their applicability to the
description of neutrino oscillations,
because any oscillation experiment involves the
production and detection of neutrinos.

It is then natural to ask if the standard theory of neutrino oscillations
is appropriate for the description of neutrino interaction processes.
In other words,
are the standard flavor states in Eq.~(\ref{1331})
appropriate for the calculation
of the neutrino production and detection rates?

In this paper we review the derivation the flavor neutrino states
in the framework of standard Quantum Field Theory,
clarifying some subtle issues.
We will show that the flavor neutrino states
are appropriate for the description of neutrino production and detection,
as well as for the description of neutrino oscillations\footnote{
In this review we consider neutrino oscillations on vacuum.
We do not consider the more complicated case of neutrino oscillations in matter
\cite{Wolfenstein:1978ue,Mikheev:1985gs}.
}.
We will see that
the flavor neutrino states reduce to the
standard flavor states in Eq.~(\ref{1331})
in the case of experiments which are not sensitive to the difference of the neutrino masses.
In this case,
we recover the standard oscillation probability.

In this paper we adopt the so-called ``plane wave approximation'',
in which the massive neutrino components of a flavor state are described by plane waves,
as in the standard approach
(see the reviews in
Refs.~\cite{Bilenky:1978nj,Bilenky:1987ty,hep-ph/9812360,hep-ph/0202058,hep-ph/0306239}).
However,
for the justification of some assumptions in the derivation of the
oscillation probability,
we will need to take into account the wave packet character of propagating massive neutrinos
\cite{Nussinov:1976uw,Kayser:1981ye,Giunti:1991ca,hep-ph/9506271,hep-ph/9710289,hep-ph/9711363,hep-ph/9810543,hep-ph/0205014,hep-ph/0302026,hep-ph/9305276,hep-ph/9709494,hep-ph/9812441,hep-ph/9909332,hep-ph/0109119,hep-ph/0202068},
albeit without need of a specific model.

The plan of the paper is as follows.
In section~\ref{s01}
we review the standard derivation of the neutrino oscillation probability,
highlighting the underlying assumptions.
In section~\ref{h003}
we present a derivation of the flavor neutrino states
in the framework of the standard Quantum Field Theory.
In section~\ref{h068}
we show that the flavor neutrino states are appropriate for the description of neutrino
production and detection.
In section~\ref{h024},
using these flavor states,
we derive the probability of neutrino oscillations,
taking into account the general possibility that
different massive neutrinos may have different momenta as well as different energies
\cite{Winter:1981kj,Giunti:1991ca,hep-ph/0011074,hep-ph/0104148,hep-ph/0302026}.
In section~\ref{h041} we show
that the corrections to the standard oscillation phase due
to violations of the light-ray time $=$ distance
approximation and the one-dimensional propagation approximation,
adopted in section~\ref{h024},
are negligible.
In section~\ref{301} we show that the standard oscillation probability
can also be derived through the average over the unobserved propagation time
of the space-time dependent oscillation probability,
taking into account the general properties of
the massive neutrino wave packets.

\section{Standard Derivation of the Neutrino Oscillation Probability}
\label{s01}

Neutrino oscillations are a consequence of
neutrino mixing:
\begin{equation}
\nu_{\alpha L}(x)
=
\sum_{k} U_{\alpha k} \, \nu_{kL}(x)
\qquad
(\alpha=e,\mu,\tau)
\,,
\label{001}
\end{equation}
where
$\nu_{\alpha L}(x)$
are the left-handed flavor neutrino fields,
$\nu_{kL}(x)$
are the left-handed massive neutrino fields
and
$U$ is the unitary mixing matrix
(see the reviews in
Refs.~\cite{Bilenky:1978nj,Bilenky:1987ty,hep-ph/9812360,hep-ph/0202058,hep-ph/0306239}).
Since
a flavor neutrino $\nu_{\alpha}$
is created by $\nu_{\alpha L}^{\dagger}(x)$
in a charged-current weak interaction process,
in the standard plane-wave theory of neutrino oscillations
\cite{Eliezer:1976ja,Fritzsch:1976rz,Bilenky:1976cw,Bilenky:1976yj,Bilenky:1978nj},
it is assumed that $\nu_{\alpha}$ is described
by the standard flavor state in Eq.~(\ref{1331}),
which has the same mixing as the field
$\nu_{\alpha L}^{\dagger}(x)$.

The massive neutrino states
$| \nu_k \rangle$
have definite mass $m_k$ and definite energy $E_k$.
Hence,
they evolve in time as plane waves:
\begin{equation}
i
\,
\frac{\partial}{\partial t}
\,
| \nu_k(t) \rangle
=
\mathscr{H}_{0}
\,
| \nu_k(t) \rangle
=
E_k
\,
| \nu_k(t) \rangle
\quad \Longrightarrow \quad
| \nu_k(t) \rangle
=
e^{- i E_k t}
\,
| \nu_k \rangle
\,,
\label{003}
\end{equation}
where $\mathscr{H}_{0}$ is the free Hamiltonian operator
and
$ | \nu_k(t=0) \rangle = | \nu_k \rangle $
(all the massive neutrinos start with the same arbitrary phase).
The resulting time evolution of the flavor neutrino state Eq.~(\ref{1331})
is given by
\begin{equation}
| \nu_{\alpha}(t) \rangle
=
\sum_{k} U_{\alpha k}^* \, e^{- i E_k t}
\,
| \nu_k \rangle
=
\sum_{\beta=e,\mu,\tau}
\left(
\sum_{k} U_{\alpha k}^* \, e^{- i E_k t} \, U_{\beta k}
\right)
| \nu_{\beta} \rangle
\,.
\label{004}
\end{equation}
Hence,
if the mixing matrix $U$ is different from unity
(\textit{i.e.} if there is neutrino mixing),
the state $| \nu_{\alpha}(t) \rangle$,
which has pure flavor $\alpha$ at the initial time $t=0$,
evolves in time into a superposition of different flavors.
The quantity in parentheses in Eq.~(\ref{004})
is the amplitude of $\nu_\alpha\to\nu_\beta$ transitions
at the time $t$ after $\nu_\alpha$ production.
The probability of $\nu_\alpha\to\nu_\beta$ transitions at the time $t=T$
of neutrino detection is given by
\begin{equation}
P_{\nu_\alpha\to\nu_\beta}(T)
=
|\langle \nu_{\beta} | \nu_{\alpha}(T) \rangle |^2
=
\left|
\sum_{k} U_{\alpha k}^* \, e^{- i E_k T} \, U_{\beta k}
\right|^2
=
\sum_{k,j} U_{\alpha k}^* \, U_{\beta k} \,  U_{\alpha j} \, U_{\beta j}^*
\, e^{- i ( E_k - E_j ) T}
\,.
\label{005}
\end{equation}
One can see that
$P_{\nu_\alpha\to\nu_\beta}(T)$
depends on the energy differences
$E_k - E_j$.
In the standard theory of neutrino oscillations
it is assumed that all massive neutrinos have the same momentum $\vet{p}$.
Since detectable neutrinos are ultrarelativistic\footnote{
It is known that neutrino masses
are smaller than about one eV
(see the reviews in Refs.~\cite{hep-ph/0211462,hep-ph/0310238}).
Since only neutrinos with energy larger than about 100 keV
can be detected (see the discussion in Ref.~\cite{hep-ph/0205014}),
in oscillation experiments neutrinos are always ultrarelativistic.
},
we have
\begin{equation}
E_k
=
\sqrt{\vet{p}^2+m_k^2}
\simeq
E
+
\frac{m_k^2}{2E}
\quad \Longrightarrow \quad
E_k - E_j
=
\frac{\Delta{m}^2_{kj}}{2E}
\,,
\label{006}
\end{equation}
where
$\Delta{m}^2_{kj} \equiv m_k^2 - m_j^2$
and
$E \equiv |\vet{p}|$
is the energy of a massless neutrino
(or, in other words,
the neutrino energy in the massless approximation).
In most neutrino oscillation experiments
the time $T$ between production and detection is not measured,
but the source-detector distance $L$ is known.
In this case,
in order to apply the oscillation probability to the data analysis
it is necessary to express $t$ as a function of $L$.
Considering ultrarelativistic neutrinos,
we have $T \simeq L$,
leading to the standard formula for the oscillation probability:
\begin{equation}
P_{\nu_{\alpha}\to\nu_{\beta}}(L,E)
=
\sum_{k,j}
U_{{\alpha}k}^{*}
\,
U_{{\beta}k}
\,
U_{{\alpha}j}
\,
U_{{\beta}j}^{*}
\,
\exp\left( - i \, \frac{\Delta{m}^{2}_{kj} L}{2E} \right)
\,.
\label{a001}
\end{equation}

Summarizing,
there are three main assumptions in the standard theory of neutrino oscillations:
\renewcommand{\labelenumi}{\theenumi}
\renewcommand{\theenumi}{(A\arabic{enumi})}
\begin{enumerate}
\item
\label{A1}
Neutrinos produced or detected in
charged-current weak interaction processes
are described by the
flavor states in Eq.~(\ref{1331}).
\item
\label{A2}
The massive neutrino states
$|\nu_{k}\rangle$
in Eq.~(\ref{1331})
have the same momentum
(``\emph{equal-momentum assumption}'').
\item
\label{A3}
The propagation time
is equal to the
distance $L$
traveled by the neutrino
between production and detection
(``\emph{time $=$ distance assumption}'').
\end{enumerate}
In the following we will show that the assumptions~\ref{A1} and ~\ref{A3}
correspond to approximations which are appropriate in the analysis of current
neutrino oscillation experiments.
Instead,
the equal-momentum assumption~\ref{A2} is not physically justified
\cite{Winter:1981kj,Giunti:1991ca,hep-ph/0011074,hep-ph/0104148,hep-ph/0302026},
as one can easily understand from the application of energy-momentum conservation
to the production process\footnote{
A different opinion,
in favor of the equal-momentum assumption,
has been recently expressed in
Ref.~\cite{hep-ph/0604044}.
On the other hand,
other authors \cite{hep-ph/9607201,hep-ph/9802387,hep-ph/0304187}
advocated an equal-energy assumption,
which we consider as unphysical as the equal-momentum assumption.
}.
However,
in section~\ref{h024} we will show that the assumption~\ref{A2}
is actually not necessary for the derivation of the
oscillation probability if both the evolutions in space and in time
of the neutrino state are taken into account.

\section{Flavor Neutrino States}
\label{h003}

The state of a flavor neutrino $\nu_{\alpha}$
is defined as the state
which describes a neutrino produced in a charged-current weak interaction process
together with a charged lepton
$\ell_{\alpha}^{+}$
or from a charged lepton
$\ell_{\alpha}^{-}$
($ \ell_{\alpha}^{\pm} = e^{\pm} , \mu^{\pm} , \tau^{\pm} $ for $\alpha=e,\mu,\tau$, respectively),
or the state
which describes a neutrino detected in a charged-current weak interaction process
with a charged lepton
$\ell_{\alpha}^{-}$ in the final state.
In fact, the neutrino flavor can only be measured
through the identification of the charged lepton
associated with the neutrino
in a charged-current weak interaction process.

Let us first consider a neutrino produced in the generic decay process
\begin{equation}
\text{P}_{\text{I}} \to \text{P}_{\text{F}} + \ell_{\alpha}^{+} + \nu_{\alpha}
\,,
\label{h007}
\end{equation}
where $\text{P}_{\text{I}}$ is the decaying particle
and $\text{P}_{\text{F}}$ represents any number of final particles.
For example: in the pion decay process
\begin{equation}
\pi^{+} \to \mu^{+} + \nu_{\mu}
\,,
\label{h008}
\end{equation}
we have $\text{P}_{\text{I}}=\pi^{+}$, $\text{P}_{\text{F}}$ is absent and $\alpha=\mu$;
in a nuclear $\beta^{+}$ decay process
$
\text{N}(A,Z)
\to
\text{N}(A,Z-1)
+
e^{+} + \nu_{e}
$
we have $\text{P}_{\text{I}}=\text{N}(A,Z)$, $\text{P}_{\text{F}}=\text{N}(A,Z-1)$ and $\alpha=e$.
The following method can easily be
modified in the case of a $\nu_{\alpha}$ produced in the generic scattering process
$ \ell_{\alpha}^{-} + \text{P}_{\text{I}} \to \text{P}_{\text{F}} + \nu_{\alpha} $
by replacing the $\ell_{\alpha}^{+}$ in the final state with a $\ell_{\alpha}^{-}$
in the initial state.

The final state resulting from the decay
of the initial particle $\text{P}_{\text{I}}$ is given by
\begin{equation}
| f \rangle
=
\mathsf{S} \, | \text{P}_{\text{I}} \rangle
\,,
\label{h011}
\end{equation}
where $\mathsf{S}$
is the $S$-matrix operator.
Since
the final state $ | f \rangle $ contains all the decay channels of $\text{P}_{\text{I}}$,
it can be written as
\begin{equation}
| f \rangle
=
\sum_{k} \mathcal{A}^{\text{P}}_{\alpha k} \, | \nu_{k} , \ell_{\alpha}^{+} , \text{P}_{\text{F}} \rangle
+
\ldots
\,,
\label{h012}
\end{equation}
where we have singled out the decay channel in Eq.~(\ref{h007})
and we have taken into account that the flavor neutrino $\nu_{\alpha}$
is a coherent superposition of massive neutrinos $\nu_{k}$.
Since
the states of the other decay channels represented by dots in Eq.~(\ref{h012})
are orthogonal to $| \nu_{k} , \ell_{\alpha}^{+} , \text{P}_{\text{F}} \rangle$
and
the different states
$| \nu_{k} , \ell_{\alpha}^{+} , \text{P}_{\text{F}} \rangle$
are orthogonal and normalized,
the coefficients $\mathcal{A}^{\text{P}}_{\alpha k}$
are the amplitudes of production of the corresponding state
in the decay channel in Eq.~(\ref{h007}):
\begin{equation}
\mathcal{A}^{\text{P}}_{\alpha k}
=
\langle \nu_{k} , \ell_{\alpha}^{+} , \text{P}_{\text{F}} | f \rangle
=
\langle \nu_{k} , \ell_{\alpha}^{+} , \text{P}_{\text{F}} |
\,
\mathsf{S}
\,
| \text{P}_{\text{I}} \rangle
\,.
\label{h013}
\end{equation}
Projecting the final state in Eq.~(\ref{h012})
over $|\ell_{\alpha}^{+} , \text{P}_{\text{F}} \rangle$
and normalizing,
we obtain the flavor neutrino state
\cite{Giunti:1992cb,hep-ph/0102320,hep-ph/0306239,hep-ph/0402217}
\begin{equation}
| \nu_{\alpha}^{\text{P}} \rangle
=
\left( \sum_{i} |\mathcal{A}^{\text{P}}_{\alpha i}|^{2} \right)^{-1/2}
\sum_{k} \mathcal{A}^{\text{P}}_{\alpha k} \, | \nu_{k} \rangle
\,.
\label{h015}
\end{equation}
Therefore, a flavor neutrino state is a coherent superposition of massive neutrino states
$ | \nu_{k} \rangle $
and
the coefficient
$\mathcal{A}^{\text{P}}_{\alpha k}$
of the $k^{\text{th}}$ massive neutrino component
is given by the amplitude of production of $\nu_{k}$.
Since, in general,
the amplitudes $\mathcal{A}^{\text{P}}_{\alpha k}$ depend on the production process,
a flavor neutrino state depends on the production process.
In the following,
we will call a flavor neutrino state of the type in Eq.~(\ref{h015}) a
``\emph{production flavor neutrino state}''.

Let us now consider the detection of a flavor neutrino $\nu_{\alpha}$
through the generic charged-current weak interaction process
\begin{equation}
\nu_{\alpha} + \text{D}_{\text{I}} \to \text{D}_{\text{F}} + \ell_{\alpha}^{-}
\,,
\label{h035}
\end{equation}
where $\text{D}_{\text{I}}$ is the target particle
and $\text{D}_{\text{F}}$ represents one or more final particles.
In general,
since the incoming neutrino state in the detection process is a superposition of
massive neutrino states, it may not have a definite flavor.
Therefore,
we must consider the generic process
\begin{equation}
\nu + \text{D}_{\text{I}}
\,,
\label{h035a}
\end{equation}
with a generic incoming neutrino state $ | \nu \rangle $.
In this case,
the final state of the scattering process is given by
\begin{equation}
| f \rangle
=
\mathsf{S} \, | \nu , \text{D}_{\text{I}} \rangle
\,,
\label{h201}
\end{equation}
This final state contains all the possible scattering channels:
\begin{equation}
| f \rangle
=
| \text{D}_{\text{F}} , \ell_{\alpha}^{-} \rangle + \ldots
\,,
\label{h201a}
\end{equation}
where we have singled out the scattering channel in Eq.~(\ref{h035}).
We want to find the component
\begin{equation}
| \nu_{\alpha} , \text{D}_{\text{I}} \rangle
=
\sum_{k} \mathcal{A}^{\text{D}}_{\alpha k} | \nu_{k} , \text{D}_{\text{I}} \rangle
\label{h203}
\end{equation}
of the initial state
$ | \nu , \text{D}_{\text{I}} \rangle $
which corresponds to the flavor $\alpha$,
i.e.\ the component which generates only the scattering channel in Eq.~(\ref{h035}).
This means that
$
| \text{D}_{\text{F}} , \ell_{\alpha}^{-} \rangle
=
\mathsf{S} \, | \nu_{\alpha} , \text{D}_{\text{I}} \rangle
$.
Using the unitarity of the mixing matrix,
we obtain
\begin{equation}
| \nu_{\alpha} , \text{D}_{\text{I}} \rangle
=
\mathsf{S}^{\dagger} \, | \text{D}_{\text{F}} , \ell_{\alpha}^{-} \rangle
\,.
\label{h202}
\end{equation}
From Eqs.~(\ref{h203}) and (\ref{h202}),
the coefficients $\mathcal{A}^{\text{D}}_{\alpha k}$ are the complex conjugate of the
amplitude of detection of $\nu_{k}$ in the detection process in Eq.~(\ref{h035}):
\begin{equation}
\mathcal{A}^{\text{D}}_{\alpha k}
=
\langle \nu_{k} , \text{D}_{\text{I}} | \mathsf{S}^{\dagger} \, | \text{D}_{\text{F}} , \ell_{\alpha}^{-} \rangle
\,.
\label{h204}
\end{equation}
Projecting $ | \nu_{\alpha} , \text{D}_{\text{I}} \rangle $ over $ | \text{D}_{\text{I}} \rangle $ and normalizing,
we finally obtain the flavor neutrino state in the detection process in Eq.~(\ref{h035}):
\begin{equation}
| \nu_{\alpha}^{\text{D}} \rangle
=
\left( \sum_{i} |\mathcal{A}^{\text{D}}_{\alpha i}|^{2} \right)^{-1/2}
\sum_{k} \mathcal{A}^{\text{D}}_{\alpha k} \, | \nu_{k} \rangle
\,.
\label{h205}
\end{equation}
In the following,
we will call a flavor neutrino state of this type a
``\emph{detection flavor neutrino state}''.

Although
the expressions in Eqs.~(\ref{h015}) and (\ref{h205})
for the production and detection flavor neutrino states
have the same structure,
these states have different meanings.
A production flavor neutrino state describes the neutrino created in a
charged-current interaction process, which propagates out of a source.
Hence, it describes the initial state of a propagating neutrino.
A detection flavor neutrino state does not describe a propagating neutrino.
It describes the component of the state of a propagating neutrino which
can generate a charged lepton with appropriate flavor
through a charged-current weak interaction with
an appropriate target particle.
In other words, the scalar product
\begin{equation}
A_{\alpha}
=
\langle \nu_{\alpha}^{\text{D}} | \nu \rangle
\label{h206}
\end{equation}
is the probability amplitude to find a $\nu_{\alpha}$
by observing the scattering channel in Eq.~(\ref{h035}) with the scattering process in Eq.~(\ref{h035a}).

In order to understand the connection of the production and detection flavor neutrino states
with the standard flavor neutrino states in Eq.~(\ref{1331}),
it is useful to express the $S$-matrix operator as
\begin{equation}
\mathsf{S}
=
1
- i \int \text{d}^{4}x \,
\mathscr{H}_{\text{CC}}(x)
\,,
\label{h016}
\end{equation}
where we have considered
only the first order perturbative contribution of the
effective low-energy charged-current weak interaction Hamiltonian
\begin{equation}
\mathscr{H}_{\text{CC}}(x)
=
\frac{ G_{\text{F}} }{ \sqrt{2} }
\,
j_{\rho}^{\dagger}(x) \, j^{\rho}(x)
\,,
\label{h016a}
\end{equation}
where $G_{\text{F}}$ is the Fermi constant.
The weak charged current $j^{\rho}(x)$ is given by
\begin{align}
j^{\rho}(x)
=
\null & \null
\sum_{\alpha=e,\mu,\tau}
\overline{\nu_{\alpha}}(x)
\,
\gamma^{\rho}
\left( 1 - \gamma^{5} \right)
\ell_{\alpha}(x)
+
h^{\rho}(x)
\nonumber
\\
=
\null & \null
\sum_{\alpha=e,\mu,\tau}
\sum_{k}
U_{\alpha k}^{*}
\,
\overline{\nu_{k}}(x)
\,
\gamma^{\rho}
\left( 1 - \gamma^{5} \right)
\ell_{\alpha}(x)
+
h^{\rho}(x)
\,,
\label{h017}
\end{align}
where $h^{\rho}(x)$ is the hadronic weak charged current.
The production and detection amplitudes
$\mathcal{A}^{\text{P}}_{\alpha k}$
and
$\mathcal{A}^{\text{D}}_{\alpha k}$
can be written as
\begin{equation}
\mathcal{A}^{\text{P}}_{\alpha k}
=
U_{\alpha k}^{*}
\,
\mathcal{M}^{\text{P}}_{\alpha k}
\,,
\qquad
\mathcal{A}^{\text{D}}_{\alpha k}
=
U_{\alpha k}^{*}
\,
\mathcal{M}^{\text{D}}_{\alpha k}
\,,
\label{h018}
\end{equation}
with the interaction matrix elements
\begin{align}
\null & \null
\mathcal{M}^{\text{P}}_{\alpha k}
=
- i
\,
\frac{G_{\text{F}}}{\sqrt{2}}
\int \text{d}^{4}x \,
\langle \nu_{k} , \ell_{\alpha}^{+} |
\,
\overline{\nu_{k}}(x)
\,
\gamma^{\rho}
\left( 1 - \gamma^{5} \right)
\ell_{\alpha}(x)
\,
| 0 \rangle
\,
J_{\rho}^{\text{P}_{\text{I}} \to \text{P}_{\text{F}}}(x)
\,,
\label{h019}
\\
\null & \null
\mathcal{M}^{\text{D}}_{\alpha k}
=
i
\,
\frac{G_{\text{F}}}{\sqrt{2}}
\int \text{d}^{4}x \,
\langle \nu_{k} |
\,
\overline{\nu_{k}}(x)
\,
\gamma^{\rho}
\left( 1 - \gamma^{5} \right)
\ell_{\alpha}(x)
\,
| \ell_{\alpha}^{-} \rangle
\,
{J_{\rho}^{\text{D}_{\text{I}} \to \text{D}_{\text{F}}}}^{*}(x)
\,.
\label{h036}
\end{align}
Here
$J_{\rho}^{\text{P}_{\text{I}} \to \text{P}_{\text{F}}}(x)$
and
$J_{\rho}^{\text{D}_{\text{I}} \to \text{D}_{\text{F}}}(x)$
are,
respectively,
the matrix elements of the
$\text{P}_{\text{I}} \to \text{P}_{\text{F}}$
and
$\text{D}_{\text{I}} \to \text{D}_{\text{F}}$
transitions.

Using Eq.~(\ref{h018}),
the production and detection flavor neutrino states
can be written as
\begin{align}
\null & \null
| \nu_{\alpha}^{\text{P}} \rangle
=
\sum_{k}
\frac{\mathcal{M}^{\text{P}}_{\alpha k}}{ \sqrt{ \sum_{j} |U_{\alpha j}|^{2} \, |\mathcal{M}^{\text{P}}_{\alpha j}|^{2} } }
\,
U_{\alpha k}^{*}
\,
| \nu_{k} \rangle
\,,
\label{h020a}
\\
\null & \null
| \nu_{\alpha}^{\text{D}} \rangle
=
\sum_{k}
\frac{\mathcal{M}^{\text{D}}_{\alpha k}}{ \sqrt{ \sum_{j} |U_{\alpha j}|^{2} \, |\mathcal{M}^{\text{D}}_{\alpha j}|^{2} } }
\,
U_{\alpha k}^{*}
\,
| \nu_{k} \rangle
\,.
\label{h020b}
\end{align}
These states have a structure
which is similar to the standard flavor states in Eq.~(\ref{1331}),
with the relative contribution of
the massive neutrino $\nu_{k}$ proportional to $U_{\alpha k}^{*}$.
The additional factors
are due to the dependence of
the production and detection processes on the neutrino masses.

In experiments which are not sensitive to the dependence of
$\mathcal{M}^{\text{P}}_{\alpha k}$
or
$\mathcal{M}^{\text{D}}_{\alpha k}$
on the difference of the neutrino masses
it is possible to approximate
\begin{equation}
\mathcal{M}^{\text{P}}_{\alpha k} \simeq \mathcal{M}^{\text{P}}_{\alpha}
\,,
\qquad
\mathcal{M}^{\text{D}}_{\alpha k} \simeq \mathcal{M}^{\text{D}}_{\alpha}
\,.
\label{h021}
\end{equation}
In this case, since
\begin{equation}
\displaystyle \sum_{k} |U_{\alpha k}|^{2} = 1
\,,
\label{h022}
\end{equation}
we obtain,
up to an irrelevant phase,
the standard flavor neutrino states in Eq.~(\ref{1331}),
which do not depend on the production or detection process.
Hence,
the standard flavor neutrino states
are approximations of the production and detection flavor neutrino states
in experiments which are not sensitive to the dependence of
the neutrino interaction rate
on the difference of the neutrino masses.

In the following section~\ref{h068}
we will show that the correct expressions for the production and detection flavor neutrino states
are important in order to be able to describe, in a consistent framework,
neutrino oscillations
and
neutrino production and detection.
Then, in the next section~\ref{h024}
we will derive the neutrino oscillation probability
starting from the production and detection flavor neutrino states.
We will show that, with the appropriate approximations,
the oscillation probability
reduces to the standard one in Eq.~(\ref{a001}),
which is derived from the
approximate flavor neutrino states in Eq.~(\ref{1331}).

\section{Production and Detection Rates}
\label{h068}

In order to measure
$\nu_{\alpha}\to\nu_{\beta}$
oscillations, it is necessary to
calculate the neutrino production rate $\Gamma_{\alpha}(E)$ of $\nu_{\alpha}$ in the source
and the detection cross section $\sigma_{\beta}(E)$ of $\nu_{\beta}$.
The number of transition events as a function of the distance $L$
traveled by the neutrino between production and detection
and the neutrino energy $E$
is given by
\begin{equation}
N_{\alpha\beta}(L,E)
\propto
\Gamma_{\alpha}(E)
\,
P_{\nu_{\alpha}\to\nu_{\beta}}(L,E)
\,
\sigma_{\beta}(E)
\,,
\label{h069}
\end{equation}
with a constant of proportionality which depends on the size and composition
of the source and detector
and on the running time of the experiment.
From the measurement of $N_{\alpha\beta}(L,E)$
and the knowledge of
$\Gamma_{\alpha}(E)$ and $\sigma_{\beta}(E)$,
the experimentalist infers the value of
$P_{\nu_{\alpha}\to\nu_{\beta}}(L,E)$,
which gives information on the mixing parameters
(elements of the mixing matrix and squared-mass differences)
through Eq.~(\ref{a001}).

Decay rates and cross sections
are given by the incoherent sum over the different channels
corresponding to different massive neutrinos
\cite{Shrock:1980vy,McKellar:1980cn,Kobzarev:1980nk,Shrock:1981ct,Shrock:1981wq}.
The reason is that massive neutrinos
are the physical particles
which propagate in space-time with definite kinematical properties.
However,
in section~\ref{h024}
we derived the oscillation probability $P_{\nu_{\alpha}\to\nu_{\beta}}(L,E)$
starting from the description of neutrinos through flavor states,
which are coherent superpositions of massive neutrino states.
Then it is natural to ask if the derivation of
the oscillation probability is consistent with the
calculation of decay rates and cross sections.
In the following, we show that
the description of neutrinos through flavor states
leads to the correct expression for the
decay rates and cross sections.
Therefore,
all quantities in Eq.~(\ref{h069})
can be calculated in a consistent way.

In order to be definite,
we consider the general decay process in Eq.~(\ref{h007}),
in which a flavor neutrino $\nu_{\alpha}$ is produced.
Any other process of neutrino production or detection can be treated in an analogous way.

Using the flavor state in Eq.~(\ref{h015}) and taking into account Eq.~(\ref{h013}),
the amplitude of the general decay process in Eq.~(\ref{h007})
is given by \cite{hep-ph/0402217}
\begin{equation}
\mathcal{A}^{\text{P}}_{\alpha}
=
\langle \nu_{\alpha}^{\text{P}} , \ell_{\alpha}^{+} , \text{P}_{\text{F}} |
\,
\mathsf{S}
\,
| \text{P}_{\text{I}} \rangle
=
\left( \sum_{i} |\mathcal{A}^{\text{P}}_{\alpha i}|^{2} \right)^{-1/2}
\sum_{k} \mathcal{A}^{\text{P}*}_{\alpha k}
\,
\langle \nu_{k} , \ell_{\alpha}^{+} , \text{P}_{\text{F}} |
\mathsf{S}
| \text{P}_{\text{I}} \rangle
=
\sqrt{ \sum_{i} |\mathcal{A}^{\text{P}}_{\alpha i}|^{2} }
\,.
\label{h070}
\end{equation}
Therefore,
the decay probability is correctly given by an incoherent
sum of the probabilities of production of different massive neutrinos,
\begin{equation}
|\mathcal{A}^{\text{P}}_{\alpha}|^{2}
=
\sum_{i} |\mathcal{A}^{\text{P}}_{\alpha i}|^{2}
\,.
\label{h071}
\end{equation}
In other words,
the coherent character of the flavor state in Eq.~(\ref{h015})
is irrelevant for the decay probability,
which can be obtained
either using the flavor neutrino state in Eq.~(\ref{h015})
or
an incoherent mixture of massive neutrino states.
The decay rate is then obtained by integrating each massive neutrino contribution
to the decay probability over its phase space.

Using the expression in Eq.~(\ref{h018}) for the amplitude $\mathcal{A}^{\text{P}}_{\alpha k}$,
the decay probability in Eq.~(\ref{h071}) can be written as
\begin{equation}
|\mathcal{A}^{\text{P}}_{\alpha}|^{2}
=
\sum_{k} |U_{\alpha k}|^{2} \, |\mathcal{M}^{\text{P}}_{\alpha k}|^{2}
\,,
\label{h072}
\end{equation}
which is
an incoherent
sum of the probabilities of production of the different massive neutrinos
weighted by $|U_{\alpha k}|^{2}$
\cite{Shrock:1980vy,McKellar:1980cn,Kobzarev:1980nk,Shrock:1981ct,Shrock:1981wq}.

Therefore,
the flavor neutrino state in Eq.~(\ref{h015})
leads to the correct decay rate for the general decay process in Eq.~(\ref{h007}).
It is clear that
this proof can easily be generalized to any charged-current
weak interaction process in which flavor neutrinos
are produced or detected.

If an experiment is not sensitive to the dependence of
$\mathcal{M}^{\text{P}}_{\alpha k}$
on the different neutrino masses,
it is possible to use the approximation in Eq.~(\ref{h021}).
In this case, using Eq.~(\ref{h022}),
we obtain
\begin{equation}
|\mathcal{A}^{\text{P}}_{\alpha}|^{2}
=
|\mathcal{M}^{\text{P}}_{\alpha}|^{2}
\,.
\label{h073}
\end{equation}
If the scale of neutrino masses is negligible
in comparison with the experimental resolution,
the dependence of $\mathcal{M}^{\text{P}}_{\alpha}$
on the neutrino masses is negligible and
the decay probability in Eq.~(\ref{h073})
reduces to the standard decay probability for massless neutrinos.

The decay probability in Eq.~(\ref{h073})
can also be obtained starting from the standard flavor states in Eq.~(\ref{1331}),
which are
obtained from Eq.~(\ref{h015}) through the approximation in Eq.~(\ref{h021}).
Indeed,
in this case the decay amplitude is given by \cite{hep-ph/0302045}
\begin{equation}
\mathcal{A}^{\text{P}}_{\alpha}
=
\langle \nu_{\alpha} , \ell_{\alpha}^{+} , \text{P}_{\text{F}} |
\,
\mathsf{S}
\,
| \text{P}_{\text{I}} \rangle
=
\sum_{k} U_{\alpha k} \, \mathcal{A}^{\text{P}}_{\alpha k}
=
\sum_{k} |U_{\alpha k}|^{2} \, \mathcal{M}^{\text{P}}_{\alpha}
=
\mathcal{M}^{\text{P}}_{\alpha}
\,.
\label{h074}
\end{equation}
Let us remark, however,
that in the case of
an experiment which is sensitive to the dependence of $\mathcal{M}^{\text{P}}_{\alpha k}$
on the different neutrino masses,
a derivation of the decay amplitude starting from the standard flavor states
would lead to an incorrect result.
This is due to the fact that in this case the approximation in Eq.~(\ref{h021})
is not valid and one must take into account the dependence of $\mathcal{M}^{\text{P}}_{\alpha k}$
on the different neutrino masses in the definition of the
flavor states.

\section{Neutrino Oscillations}
\label{h024}

Let us consider a neutrino oscillation experiment
in which $ \nu_{\alpha} \to \nu_{\beta} $
transitions are studied
with a production process of the type in Eq.~(\ref{h007})
and a detection process of the type in Eq.~(\ref{h035}).
In this case,
the produced flavor neutrino $ \nu_{\alpha} $
is described by the production flavor state $ | \nu_{\alpha}^{\text{P}} \rangle $ in Eq.~(\ref{h015}).
If the neutrino production and detection processes
are separated by a space-time interval
$(\vet{L},T)$,
the neutrino propagates freely between production and detection,
evolving into the state
\begin{equation}
| \nu (\vet{L},T) \rangle
=
e^{ -i \mathsf{P}^{0} T + i \vet{\mathsf{P}} \cdot \vet{L} }
\,
| \nu_{\alpha}^{\text{P}} \rangle
\,,
\label{h027a}
\end{equation}
where
$\mathsf{P}^{0}$ and $\vet{\mathsf{P}}$
are, respectively, the energy and momentum operators.
This is the incoming neutrino state in the detection process.
The amplitude of the measurable $ \nu_{\alpha} \to \nu_{\beta} $ transitions
is given by the scalar product in Eq.~(\ref{h206}):
\begin{equation}
A_{\nu_{\alpha}\to\nu_{\beta}}(\vet{L},T)
=
\langle \nu_{\beta}^{\text{D}} | \nu (\vet{L},T) \rangle
=
\langle \nu_{\beta}^{\text{D}} |
e^{ -i \mathsf{P}^{0} T + i \vet{\mathsf{P}} \cdot \vet{L} }
| \nu_{\alpha}^{\text{P}} \rangle
\,,
\label{h027}
\end{equation}
with the detection flavor state
$ | \nu_{\beta}^{\text{D}} \rangle $
in Eq.~(\ref{h205}).

Since the massive neutrinos have definite kinematical properties
(energy and momentum),
we have,
in the plane wave approximation,
\begin{equation}
\mathsf{P}^{\mu} \, |\nu_{k}\rangle
=
p_{k}^{\mu} \, |\nu_{k}\rangle
\,,
\label{h028}
\end{equation}
with
\begin{equation}
p_{k}^{0} = E_{k} = \sqrt{ |\vet{p}_{k}|^{2} + m_{k}^{2} }
\,.
\label{h029}
\end{equation}
Using the normalization
$ \langle \nu_{k} | \nu_{j} \rangle = \delta_{kj} $,
we obtain the flavor transition amplitude
\begin{equation}
A_{\nu_{\alpha}\to\nu_{\beta}}(\vet{L},T)
=
\left( \sum_{i} |\mathcal{A}^{\text{P}}_{\alpha i}|^{2} \right)^{-1/2}
\left( \sum_{i} |\mathcal{A}^{\text{D}}_{\beta i}|^{2} \right)^{-1/2}
\sum_{k}
\mathcal{A}^{\text{P}}_{\alpha k}
\,
\mathcal{A}^{\text{D}*}_{\beta k}
\,
e^{ - i E_{k} T + i \vet{p}_{k} \cdot \vet{L} }
\,,
\label{h030}
\end{equation}
Notice that the consideration of
the space-time interval
between neutrino production and detection allows one to
take into account both the differences
in energy and momentum of massive neutrinos
\cite{Winter:1981kj,Giunti:1991ca,hep-ph/0011074,hep-ph/0104148,hep-ph/0302026}.

Let us consider the simplest case in which
all massive neutrino momenta $\vet{p}_{k}$ are aligned along $\vet{L}$.
This assumption is reasonable,
because
all massive neutrino components
are created in the same microscopic production process
and detected in the same microscopic detection process,
after propagation through the large macroscopic space interval $\vet{L}$.
In section~\ref{h041}
we will show that possible deviations
from this assumption do not lead to any observable effect.

In this ``one-dimensional approximation'',
the transition amplitude depends on $L\equiv|\vet{L}|$ and $T$:
\begin{equation}
A_{\nu_{\alpha}\to\nu_{\beta}}(L,T)
=
\left( \sum_{i} |\mathcal{A}^{\text{P}}_{\alpha i}|^{2} \right)^{-1/2}
\left( \sum_{i} |\mathcal{A}^{\text{D}}_{\beta i}|^{2} \right)^{-1/2}
\sum_{k}
\mathcal{A}^{\text{P}}_{\alpha k}
\,
\mathcal{A}^{\text{D}*}_{\beta k}
\,
e^{ - i E_{k} T + i p_{k} L }
\,,
\label{h031}
\end{equation}
where
$p_{k}\equiv|\vet{p}_{k}|$.

In oscillation experiments in which
the neutrino propagation time $T$ is not measured,
it is possible to adopt the light-ray $T = L$ approximation,
since neutrinos are ultrarelativistic
(the effects of possible deviations from $T = L$
are shown to be negligible in section~\ref{h041}).
In this case,
the phase in Eq.~(\ref{h031}) becomes
\begin{equation}
- E_{k} T + p_{k} L
=
- \left( E_{k} - p_{k} \right) L
=
- \frac{ E_{k}^{2} - p_{k}^{2} }{ E_{k} + p_{k} } \, L
=
- \frac{ m_{k}^{2} }{ E_{k} + p_{k} } \, L
\simeq
- \frac{ m_{k}^{2} }{ 2 E } \, L
\,,
\label{h032}
\end{equation}
where $E$ is the neutrino energy neglecting mass contributions.
Equation~(\ref{h032})
shows that the phases of massive neutrinos relevant for the oscillations
are independent from the values of the energies and momenta
of different massive neutrinos
\cite{Winter:1981kj,Giunti:1991ca,hep-ph/0011074,hep-ph/0104148,hep-ph/0302026},
because of the relativistic dispersion relation in Eq.~(\ref{h029}).
In particular, Eq.~(\ref{h032}) shows that
the equal-momentum assumption~\ref{A2} in section~\ref{s01},
adopted in the standard derivation of the neutrino oscillation probability,
is not necessary
in an improved derivation which takes into account
both the evolutions in space and in time
of the neutrino state.

The probability of
$ \nu_{\alpha} \to \nu_{\beta} $
transitions in space is given by
\begin{equation}
P_{\nu_{\alpha}\to\nu_{\beta}}(L,E)
=
\left( \sum_{i} |\mathcal{A}^{\text{P}}_{\alpha i}|^{2} \right)^{-1}
\left( \sum_{i} |\mathcal{A}^{\text{D}}_{\beta i}|^{2} \right)^{-1}
\sum_{k,j}
\mathcal{A}^{\text{P}}_{\alpha k}
\,
\mathcal{A}^{\text{D}*}_{\beta k}
\,
\mathcal{A}^{\text{P}*}_{\alpha j}
\,
\mathcal{A}^{\text{D}}_{\beta j}
\,
\exp\left( - i \frac{\Delta{m}^{2}_{kj}L}{2E} \right)
\,.
\label{h033}
\end{equation}
Using the decomposition in Eq.~(\ref{h018}),
the oscillation probability in Eq.~(\ref{h033})
can now be written as
\begin{align}
P_{\nu_{\alpha}\to\nu_{\beta}}(L,E)
=
\null & \null
\sum_{k,j}
\left(
\frac{ \displaystyle
\mathcal{M}^{\text{P}}_{\alpha k}
\,
\mathcal{M}^{\text{P}*}_{\alpha j}
}
{ \displaystyle
\sum_{i} |U_{\alpha i}|^{2} |\mathcal{M}^{\text{P}}_{\alpha i}|^{2}
}
\right)
\left(
\frac{ \displaystyle
\mathcal{M}^{\text{D}*}_{\beta k}
\,
\mathcal{M}^{\text{D}}_{\beta j}
}
{ \displaystyle
\sum_{i} |U_{\beta i}|^{2} |\mathcal{M}^{\text{D}}_{\beta i}|^{2}
}
\right)
\nonumber
\\
\null & \null
\hspace{3cm}
\times
U_{{\alpha}k}^{*}
\,
U_{{\beta}k}
\,
U_{{\alpha}j}
\,
U_{{\beta}j}^{*}
\,
\exp\left( - i \frac{\Delta{m}^{2}_{kj}L}{2E} \right)
\,.
\label{h037}
\end{align}
This probability has the same structure as the standard oscillation probability in Eq.~(\ref{a001}),
with additional factors that take into account
the effect of the neutrino masses in the production and detection processes.
It is clear from Eq.~(\ref{h037})
that these effects have an influence on the amplitude of the oscillations,
but not on the phase, which coincides with the standard one in Eq.~(\ref{a001}).

Since neutrinos in oscillation experiments are ultra-relativistic
and the experiments are not sensitive to
the dependence of neutrino interactions on the neutrino masses,
the dependence of
$\mathcal{M}^{\text{P}}_{\alpha k}$
and
$\mathcal{M}^{\text{D}}_{\beta k}$
on the neutrino masses can be neglected,
leading to the approximation in Eq.~(\ref{h021}).
In this case,
the transition probability in Eq.~(\ref{h037})
reduces to the standard one in Eq.~(\ref{a001}),
which can be obtained starting from the standard flavor states
in Eq.~(\ref{1331}).
As shown in section~\ref{h003},
the standard flavor states are obtained from the production and detection flavor states
under the same approximations.
Therefore,
the standard flavor states
are appropriate for the description of neutrino oscillation experiments
in the plane wave approximation,
as long as the
dependence of the production and detection probabilities on the neutrino masses
is negligible.
These considerations justify the assumption~\ref{A1} in section~\ref{s01},
adopted in the standard derivation of the neutrino oscillation probability.

\section{Universality of the Oscillation Phases}
\label{h041}

In the previous section~\ref{h024},
we derived the probability of neutrino oscillations
under the assumptions that
$T = L$
and
all massive neutrino momenta $\vet{p}_{k}$ are aligned along $\vet{L}$.
In this section we will show that
possible deviations from these assumptions do not affect in a significant way the
oscillation phases measured in neutrino oscillation experiments,
which are correctly given by the standard expression in Eq.~(\ref{a001}).

The momentum of a massive neutrino created in a production process
depends on the characteristics of the interaction,
on the nature of the other particles taking part in the process
and on the neutrino mass.
Let us call $\vet{p}$ and $E=|\vet{p}|$, respectively,
the momentum and energy of a massless neutrino.
The first order contribution of the neutrino mass to $\vet{p}_{k}$ and $E_{k}$
must be proportional to $m_{k}^{2}$,
because the energy-momentum dispersion relation in Eq.~(\ref{h029})
depends on $m_{k}^{2}$.
Therefore,
in general, the momentum $\vet{p}_{k}$
can be written to first order in $m_{k}^{2}$ as
\begin{equation}
\vet{p}_{k}
\simeq
\vet{p}
-
\vet{\xi}
\,
\frac{m_{k}^{2}}{2E}
\,,
\label{h044}
\end{equation}
where
\begin{equation}
\frac{\vet{\xi}}{2E}
=
-
\left.
\frac
{\partial\vet{p}_{k}}
{\partial m_{k}^{2}}
\right|_{m_{k}=0}
\,.
\label{h045}
\end{equation}
The value of the vector $\vet{\xi}$ depends on the production process.
However,
as we will see in the following,
the measurable oscillation phases are independent of $\vet{\xi}$.
Therefore, they are universal,
i.e.\ independent of the specific nature of
the neutrino production process,
as well as detection.

In the approximation in Eq.~(\ref{h044}),
the dispersion relation in Eq.~(\ref{h029}) implies that the energy of $\nu_{k}$ is given by
\begin{equation}
E_{k}
\simeq
E
+
\left(
1
-
\frac{\vet{p}\cdot\vet{\xi}}{E}
\right)
\frac{m_{k}^{2}}{2E}
\,.
\label{h046}
\end{equation}

Note that Eq.~(\ref{h044}) implies that the directions of propagation of
different massive neutrinos are slightly different if $\vet{\xi}$ is not collinear with $\vet{p}$.
How this is possible can be illustrated in the case of
the pion decay production process in Eq.~(\ref{h008}) as follows.
In the rest-frame of the pion,
the energy of $\nu_{k}$ can be calculated
from energy-momentum conservation to be given by
\begin{equation}
E_{k}
=
\frac{ m_{\pi} }{ 2 }
\left( 1 - \frac{ m_{\mu}^2 }{ m_{\pi}^2 } \right)
+
\frac{ m_{k}^2 }{ 2 \, m_{\pi} }
\,,
\label{0031}
\end{equation}
where
$m_{\pi}$
and
$m_{\mu}$
are the masses
of the pion and muon, respectively.
Since, in the rest-frame of the pion,
the neutrino and the muon are emitted back-to back,
all the massive neutrinos are emitted in the same direction.
Thus, $\vet{\xi}$ is collinear with $\vet{p}$, leading to
\begin{equation}
E_{k}
=
E
+
\left(
1
-
\xi
\right)
\frac{m_{k}^{2}}{2E}
\,,
\qquad
p_k
\equiv
|\vet{p}_{k}|
\simeq
E
-
\xi
\,
\frac{m_{k}^{2}}{2E}
\,,
\label{0033a}
\end{equation}
with
\begin{equation}
E
=
\frac{ m_{\pi} }{ 2 }
\left( 1 - \frac{ m_{\mu}^2 }{ m_{\pi}^2 } \right)
\simeq
30 \, \mathrm{MeV}
\,,
\qquad
\xi
\equiv
|\vet{\xi}|
=
\frac{1}{2}
\left( 1 + \frac{m_\mu^2}{m_\pi^2} \right)
\simeq
0.8
\,.
\label{0033}
\end{equation}
Note that the different values of the momenta and energies of
different massive neutrinos imply
different corresponding values for the momentum and energy of the outgoing muon.
Two different massive neutrinos can be produced coherently in the same decay process
only if the outgoing muon has energy and momentum uncertainties which are larger,
respectively,
of the difference of the energies and momenta of the two massive neutrinos.
These uncertainties must come from corresponding uncertainties for the pion.
Hence,
in a rigorous treatment,
the pion and the muon must be described by wave packets
\cite{hep-ph/9305276,hep-ph/9709494,hep-ph/9810543,hep-ph/9812441,hep-ph/9909332,hep-ph/0109119,hep-ph/0202068,hep-ph/0205014}.
This implies that also the massive neutrinos must be described by wave packets
\cite{Nussinov:1976uw,Kayser:1981ye,Giunti:1991ca,hep-ph/9506271,hep-ph/9710289,hep-ph/9711363,hep-ph/9810543,hep-ph/0205014,hep-ph/0302026}.
The spatial extension of the massive neutrino wave packets
explains how it is possible that the different massive neutrinos
can be detected in the same interaction process even if their space-time trajectories are different.

Let us now consider a boosted reference frame.
If the frame is boosted in the direction of the emitted massive neutrinos,
obviously they are still collinear.
Instead, if the frame is boosted in another direction,
collinearity is lost.
For example, let us consider a frame boosted with velocity $V$ in a direction which is
orthogonal to the neutrino direction in the pion rest frame.
In this frame, we have, in an obvious notation,
\begin{equation}
p'_{k\parallel}
=
p_{k\parallel}
=
p_{k}
\,,
\qquad
p'_{k\perp}
=
- \frac{V}{\sqrt{1-V^2}} \, E_{k}
\,.
\label{1134}
\end{equation}
Hence,
the angle of propagation of $\nu_{k}$ in the boosted frame,
with respect to its direction in the rest frame,
is given by
\begin{equation}
\tan\theta'_{k}
=
\frac{ p'_{k\perp} }{ p'_{k\parallel} }
\simeq
- \frac{V}{\sqrt{1-V^2}} \left( 1 + \frac{m_{k}^{2}}{2E^{2}} \right)
\,.
\label{1135}
\end{equation}
Since this angle depends on the neutrino mass,
in the boosted frame
different massive neutrinos are seen to propagate in slightly different directions.
In fact, in this case,
$\vet{p}'$ and $\vet{\xi}'$ are not collinear,
since
\begin{equation}
p'_{\parallel}
=
p_{\parallel}
=
E
\,,
\quad
p'_{\perp}
=
- \frac{V}{\sqrt{1-V^2}} \, E
\,,
\quad
\xi'_{\parallel}
=
\xi_{\parallel}
=
\xi
\,,
\quad
\xi'_{\perp}
=
\frac{V}{\sqrt{1-V^2}} \, \left( 1 - \xi \right)
\,.
\label{1136}
\end{equation}
Since many experiments measure the
oscillations of neutrinos produced by pion decay in flight
(e.g.\ atmospheric and accelerator neutrino oscillation experiments),
in which the neutrino direction is not constrained to be collinear with the
pion direction,
in order to be realistic,
in this section
we consider the general case of not-collinear
$\vet{p}$ and $\vet{\xi}$.

Let us now calculate the phases
\begin{equation}
\phi_{k}
=
- E_{k} T + \vet{p}_{k} \cdot \vet{L}
\,,
\label{h057}
\end{equation}
in the transition amplitude in Eq.~(\ref{h030})
considering a deviation of the time $T$
between neutrino production and detection
from the light-ray approximation $T=L$:
\begin{equation}
T = \frac{L}{|\langle\vet{v}\rangle|} \left( 1 + \varepsilon_{T} \right)
\,,
\label{h047}
\end{equation}
with the average velocity
\begin{equation}
\langle\vet{v}\rangle
=
\frac{1}{N} \sum_{k=1}^{N} \frac{\vet{p}_{k}}{E_{k}}
\simeq
\frac{\vet{p}}{E}
-
\left[
\frac{\vet{p}}{E}
\left(
1
-
\frac{\vet{p}\cdot\vet{\xi}}{E}
\right)
+
\vet{\xi}
\right]
\frac{\overline{m^{2}}}{2E^{2}}
\,,
\label{h048}
\end{equation}
where $N$ is the number of massive neutrinos
and
\begin{equation}
\overline{m^{2}}
=
\frac{1}{N} \sum_{k=1}^{N} m_{k}^{2}
\label{h049}
\end{equation}
is the average of the squared neutrino masses.
The absolute value
of the average velocity is given by the usual ultra-relativistic expression
\begin{equation}
|\langle\vet{v}\rangle|
\simeq
1 - \frac{\overline{m^{2}}}{2E^{2}}
\,.
\label{h050}
\end{equation}
In Eq.~(\ref{h047}), we consider
$\varepsilon_{T} \ll 1$,
because neutrinos are ultra-relativistic
and
the deviation from $T=L/|\langle\vet{v}\rangle|$ cannot be larger than
the size of the neutrino wave packets,
which must be much smaller than $L$ in order to observe the oscillations
\cite{Kayser:1981ye,Giunti:1991ca}.
To first order in the small ratio
$\overline{m^{2}}/E^{2}$,
Eq.~(\ref{h047}) becomes
\begin{equation}
T \simeq L \left( 1 + \varepsilon_{T} \right) \left( 1 + \frac{\overline{m^{2}}}{2E^{2}} \right)
\,.
\label{h051}
\end{equation}

Now, we consider also deviations from the assumption
that
the momenta of all massive neutrinos are aligned along $\vet{L}$.
As discussed above,
the expression for the neutrino momenta in Eq.~(\ref{h044})
already implies that, in general,
the momenta of different massive neutrinos are not collinear.
Moreover, we can consider a deviation from the collinearity of
$\vet{L}$
and the average momentum
\begin{equation}
\langle\vet{p}\rangle
=
\frac{1}{N} \sum_{k=1}^{N} \vet{p}_{k}
\simeq
\vet{p}
-
\vet{\xi}
\,
\frac{\overline{m^{2}}}{2E}
\,,
\label{h052}
\end{equation}
which we can write as
\begin{equation}
\frac{\langle\vet{p}\rangle}{|\langle\vet{p}\rangle|}
=
\frac{\vet{L}}{L}
+
\vet{\varepsilon}_{L}
\,,
\label{h053}
\end{equation}
with
$|\vet{\varepsilon}_{L}| \ll 1$,
for the same reasons of $\varepsilon_{T} \ll 1$.
At zeroth order in the small ratio
$\overline{m^{2}}/E^{2}$,
we have
\begin{equation}
\vet{p}
\simeq
E
\left(
\frac{\vet{L}}{L}
+
\vet{\varepsilon}_{L}
\right)
\,,
\label{h055}
\end{equation}
which implies, from Eq.~(\ref{h046}),
\begin{equation}
E_{k}
\simeq
E
+
\left(
1
-
\frac{\vet{L}\cdot\vet{\xi}}{L}
-
\vet{\varepsilon}_{L}\cdot\vet{\xi}
\right)
\frac{m_{k}^{2}}{2E}
\,.
\label{h056}
\end{equation}

From Eqs.~(\ref{h044}), (\ref{h051}) and (\ref{h056}),
for the difference
$ \Delta\phi_{kj} = \phi_{k} - \phi_{j} $
of the phases in Eq.~(\ref{h057})
we obtain,
at first order in the small quantities
$\varepsilon_{T}$ and $|\vet{\varepsilon}_{L}|$,
\begin{equation}
\Delta\phi_{kj}
\simeq
-
\frac{ \Delta{m}^{2}_{kj} L }{ 2E }
+
\varepsilon_{kj}
\,,
\label{h059}
\end{equation}
with the contribution
\begin{equation}
\varepsilon_{kj}
=
\left[
\vet{\varepsilon}_{L}\cdot\vet{\xi}
-
\varepsilon_{T}
\left(
1 - \frac{\vet{L}\cdot\vet{\xi}}{L}
\right)
\right]
\frac{ \Delta{m}^{2}_{kj} L }{ 2E }
\label{h059a}
\end{equation}
in addition to
the standard oscillation phase in Eq.~(\ref{a001}).
However,
since
$\varepsilon_{T}\ll1$ and $|\vet{\varepsilon}_{L}|\ll1$,
the contribution
$\varepsilon_{kj}$ is non-negligible
only for $ \Delta{m}^{2}_{kj} L / 2E \gg 1 $.
But in this case oscillations are not measurable,
since they are washed out by the average over the energy resolution of the detector
(see Ref.~\cite{hep-ph/9812360}).
In the case of $ \Delta{m}^{2}_{kj} L / 2E \sim 1 $,
in which oscillations are measurable,
$\varepsilon_{kj}$ is extremely small and
can be safely neglected ($ e^{i\varepsilon_{kj}} \simeq 1 $),
leading to the validity
of the standard expression in Eq.~(\ref{a001}) for the oscillation phases.

Note the irrelevance
for $\Delta\phi_{kj}$ of the lack of collinearity
of the trajectories of different massive neutrinos if
$\varepsilon_{T}=|\vet{\varepsilon}_{L}|=0$.
It is due to the fact that the deviation from collinearity of $\nu_{k}$ and $\nu_{j}$
is proportional to $ \Delta{m}^{2}_{kj} / 2E^2 $
(see the example in Eq.~(\ref{1135})).
Thus, it induces in the phase difference $\Delta\phi_{kj}$ effects
of higher order in $ \Delta{m}^{2}_{kj} / 2E^2 $,
which are completely negligible.

In conclusion, in this section we have shown that
possible small deviations from
the light-ray approximation $T=L$
and from the collinearity of
the massive neutrino momenta $\vet{p}_{k}$ and $\vet{L}$
are irrelevant for the measurable oscillation phases,
which are universally independent from the specific characteristics of
the process of neutrino production,
as well as detection.
Hence, the time $=$ distance assumption~\ref{A3} in section~\ref{s01},
adopted in the standard derivation of the neutrino oscillation probability,
is correct.

\section{Time Average}
\label{301}

In section~\ref{h024}
we derived the probability of flavor neutrino oscillations
as a function of the source-detector distance $L$
through the light-ray $T=L$ approximation.
In the previous section~\ref{h041}
we have shown that the measurable oscillation phases are stable
against possible small deviations from the light-ray approximation.
Another way to obtain the oscillation probability in space
is through an average of the space-time dependent oscillation probability
over the unobserved propagation time $T$
\cite{Giunti:1991ca},
\begin{equation}
P_{\nu_{\alpha}\to\nu_{\beta}}(L)
\propto
\int \text{d}T \, P_{\nu_{\alpha}\to\nu_{\beta}}(L,T)
\,,
\label{301a}
\end{equation}
where the constant of proportionality must be chosen in order to satisfy the
conservation of probability constraint
\begin{equation}
\sum_{\beta} P_{\nu_{\alpha}\to\nu_{\beta}}(L)
=
\sum_{\beta} P_{\nu_{\alpha}\to\nu_{\beta}}(L,T)
=
1
\,.
\label{301b}
\end{equation}
The time average is done on the oscillation probability,
not on the amplitude,
since it is an average over the propagation times of different neutrinos in a beam,
which contribute incoherently to the oscillation probability.

Let us work, for simplicity, in the approximation in Eq.~(\ref{h021})
for the production and detection matrix elements.
From the one-dimensional approximation of the oscillation amplitude in Eq.~(\ref{h031}),
we obtain the space-time dependent oscillation probability
\begin{equation}
P_{\nu_{\alpha}\to\nu_{\beta}}(L,T)
=
\sum_{k,j}
U_{{\alpha}k}^{*}
\,
U_{{\beta}k}
\,
U_{{\alpha}j}
\,
U_{{\beta}j}^{*}
\,
e^{ - i \left( E_{k} - E_{j} \right) T + i \left( p_{k} - p_{j} \right) L }
\,.
\label{302}
\end{equation}
The integration in Eq.~(\ref{301a}) leads to
\begin{equation}
P_{\nu_{\alpha}\to\nu_{\beta}}(L)
\propto
\sum_{k,j}
U_{{\alpha}k}^{*}
\,
U_{{\beta}k}
\,
U_{{\alpha}j}
\,
U_{{\beta}j}^{*}
\,
e^{ i \left( p_{k} - p_{j} \right) L }
\,
\delta( E_{k} - E_{j} )
\,.
\label{303}
\end{equation}
Hence, the time average implies that
$ E_{k} = E_{j} $
\cite{hep-ph/0202068,hep-ph/0109119}.
However,
from Eq.~(\ref{h046}) it is clear that in the plane wave approximation
this equality is, in general, not possible.
In order to allow such an equality, it is necessary to take into account
the wave packet character of massive neutrinos,
which implies an uncertainty in energy.

In an accurate wave packet treatment,
the massive neutrinos are described by superpositions of plane waves with different momenta
\cite{Kayser:1981ye,Giunti:1991ca,hep-ph/9506271,hep-ph/9710289,hep-ph/9711363,hep-ph/9810543,hep-ph/0205014,hep-ph/0302026}.
Since we are not interested in the wave packet effects,
we adopt, for each massive neutrino $\nu_{k}$,
an approximate wave packet with a distribution in energy
$\phi(E_{k}-\widetilde{E}_{k})$,
where $\widetilde{E}_{k}$ is the average energy.
In this case,
the space-time dependent oscillation probability in Eq.~(\ref{302}) becomes
\begin{align}
P_{\nu_{\alpha}\to\nu_{\beta}}(L,T)
=
\null & \null
\sum_{k,j}
U_{{\alpha}k}^{*}
\,
U_{{\beta}k}
\,
U_{{\alpha}j}
\,
U_{{\beta}j}^{*}
\nonumber
\\
\null & \null
\times
\int \text{d}E_{k}
\int \text{d}E_{j}
\,
\phi(E_{k}-\widetilde{E}_{k})
\,
\phi^{*}(E_{j}-\widetilde{E}_{j})
\,
e^{ - i \left( E_{k} - E_{j} \right) T + i \left( p_{k} - p_{j} \right) L }
\,,
\label{304}
\end{align}
with
$ p_{k} = \sqrt{ E_{k}^{2} - m_{k}^{2} } $.
The integration over $T$ in Eq.~(\ref{301a}) leads to
\begin{equation}
P_{\nu_{\alpha}\to\nu_{\beta}}(L)
\propto
\sum_{k,j}
U_{{\alpha}k}^{*}
\,
U_{{\beta}k}
\,
U_{{\alpha}j}
\,
U_{{\beta}j}^{*}
\,
\int \text{d}E
\,
\phi(E-\widetilde{E}_{k})
\,
\phi^{*}(E-\widetilde{E}_{j})
\,
e^{ i \left( p_{k} - p_{j} \right) L }
\,,
\label{305}
\end{equation}
with
$ p_{k} = \sqrt{ E^{2} - m_{k}^{2} } $.
In the ultrarelativistic approximation
$
p_{k}
\simeq
E
-
m_{k}^{2} / 2 E
$,
we obtain
\begin{equation}
P_{\nu_{\alpha}\to\nu_{\beta}}(L)
\propto
\sum_{k,j}
U_{{\alpha}k}^{*}
\,
U_{{\beta}k}
\,
U_{{\alpha}j}
\,
U_{{\beta}j}^{*}
\,
\int \text{d}E
\,
\phi(E-\widetilde{E}_{k})
\,
\phi^{*}(E-\widetilde{E}_{j})
\,
\exp\left( - i \, \frac{\Delta{m}^{2}_{kj} L}{2E} \right)
\,.
\label{306}
\end{equation}

Now,
we assume that the energy width of the wave packets is much larger than
the difference between $\widetilde{E}_{k}$ and $\widetilde{E}_{j}$,
\begin{equation}
\phi(E-\widetilde{E}_{k})
\simeq
\phi(E-\widetilde{E}_{j})
\simeq
\phi(E-\widetilde{E})
\,,
\label{307}
\end{equation}
where $\widetilde{E}$ may be chosen as the average energy of the
wave packet of a massless neutrino
(or the average among the $\widetilde{E}_{k}$'s).
This is the condition which allows the equality
$ E_{k} = E_{j} $
without suppressing the oscillation probability.
Since the energy width of the wave packets
is inversely proportional to their spatial width
and the difference $ | \widetilde{E}_{k} - \widetilde{E}_{j} | $
is inversely proportional to the oscillation length\footnote{
Writing
$
\widetilde{E}_{k}
\simeq
\widetilde{E}
+
\left(
1
-
\xi
\right)
m_{k}^{2} / 2\widetilde{E}
$,
in analogy with Eq.~(\ref{0033a}),
we have
$
| \widetilde{E}_{k} - \widetilde{E}_{j} |
\propto
|\Delta{m}^{2}_{kj}| / 2\widetilde{E}
\propto
1 / L^{\text{osc}}_{kj}
$,
where
$
L^{\text{osc}}_{kj}
=
4 \pi \widetilde{E} / |\Delta{m}^{2}_{kj}|
$
is the oscillation length.
},
the condition in Eq.~(\ref{307}) is verified by spatial wave packets which are
much smaller than the oscillation length.
This is a necessary condition for the observation of neutrino oscillation
which is satisfied in all experiments.
Using the approximation in Eq.~(\ref{307}), the oscillation probability in Eq.~(\ref{306})
becomes
\begin{equation}
P_{\nu_{\alpha}\to\nu_{\beta}}(L)
\propto
\sum_{k,j}
U_{{\alpha}k}^{*}
\,
U_{{\beta}k}
\,
U_{{\alpha}j}
\,
U_{{\beta}j}^{*}
\,
\int \text{d}E
\,
|\phi(E-\widetilde{E})|^2
\,
\exp\left( - i \, \frac{\Delta{m}^{2}_{kj} L}{2E} \right)
\,.
\label{308}
\end{equation}

Finally, we consider a sharply peaked wave packet,
i.e.\ a wave packet
$\phi(E-\widetilde{E})$
with energy uncertainty much smaller than the average energy $\widetilde{E}$.
This is a realistic assumption,
since in practice the energy uncertainties of the
production and detection processes are much smaller
than the neutrino energy.
In this case,
the phase
$ \Delta{m}^{2}_{kj} L / 2E $
is practically constant over the size of the wave packet
when the distance $L$ is of the order of the
average oscillation length
$ 4 \pi \widetilde{E} / \Delta{m}^{2}_{kj} $.
As we have already remarked in section~\ref{h041},
this is a necessary condition for the observation of oscillations,
since for $ L \gg 4 \pi \widetilde{E} / \Delta{m}^{2}_{kj} $
the oscillations are washed out by the average over the energy resolution of the detector.
Hence,
using the normalization
$
\int \text{d}E
\,
|\phi(E-\widetilde{E})|^2
=
1
$,
we finally obtain
\begin{equation}
P_{\nu_{\alpha}\to\nu_{\beta}}(L)
=
\sum_{k,j}
U_{{\alpha}k}^{*}
\,
U_{{\beta}k}
\,
U_{{\alpha}j}
\,
U_{{\beta}j}^{*}
\,
\exp\left( - i \, \frac{\Delta{m}^{2}_{kj} L}{2\widetilde{E}} \right)
\,.
\label{309}
\end{equation}
This is the standard oscillation probability in Eq.~(\ref{a001}),
with
$\widetilde{E}=E$.

Summarizing,
in this section we have shown that
the average
of the space-time dependent oscillation probability
over the unobserved propagation time $T$
leads to
the standard oscillation probability in space,
taking into account the wave packet character
of massive neutrinos,
without need of a detailed wave packet model.

\section{Conclusions}
\label{Conclusions}

In this review
we have presented a consistent framework for the description of
neutrino oscillations and interactions,
which is appropriate for the theoretical interpretation of
neutrino oscillation experiments,
in which neutrinos are produced by a source
and detected after propagation over a macroscopic distance.
We have shown that the
flavor neutrino state that describes a neutrino produced or detected in
a charged-current weak interaction process
depends on the process under consideration
and is appropriate for the description of neutrino oscillations
as well as for the calculation of neutrino production and detection rates.
A flavor neutrino state can be approximated with
the standard expression in Eq.~(\ref{1331})
only for experiments
which are not sensitive to the
dependence of neutrino interactions on the different neutrino masses.
This is the case of all neutrino oscillation experiments.

We have also reviewed and clarified some subtle points
concerning the derivation of the oscillation probability.
In particular,
we have shown that the oscillation probability
can be derived through the usual light-ray time $=$ distance
approximation,
because possible small deviations have negligible effects
on the measurable oscillation phase.
We have also shown that the oscillation probability
can be derived in an alternative way
through an average of the space-time dependent oscillation probability
over the unobserved propagation time,
taking into account the wave packet character of massive neutrinos,
without need of a specific model.



\end{document}